# Relative relocation of earthquakes without a predefined velocity model: an example from a peculiar seismic cluster on Katla volcano's south-flank (Iceland)

Giulia Sgattoni[1,2,3*], Ólafur Guðmundsson[3], Páll Einarsson[2], Federico Lucchi[1]

[1] *Department of Biological, Geological and Environmental Sciences, University of Bologna, Bologna, Italy*

[2] *Institute of Earth Sciences, Science Institute, University of Iceland, Reykjavik, Iceland*

[3] *Department of Earth Sciences, Uppsala University, Uppsala, Sweden*

*Corresponding author:* giulia.sgattoni2@unibo.it

## Summary

Relative relocation methods are commonly used to precisely relocate earthquake clusters consisting of similar waveforms. Repeating waveforms are often recorded at volcanoes, where, however, the crust structure is expected to contain strong heterogeneities and therefore the 1D velocity model assumption that is made in most location strategies is not likely to describe reality. A peculiar cluster of repeating low-frequency seismic events was recorded on the south flank of Katla volcano (Iceland) from 2011. As the hypocentres are located at the rim of the glacier, the seismicity may be due to volcanic or glacial processes. Information on the size and shape of the cluster may help constraining the source process. The extreme similarity of waveforms points to a very small spatial distribution of hypocentres. In order to extract meaningful information about size and shape of the cluster, we minimize uncertainty by optimizing the cross-correlation measurements and relative-relocation process. With a synthetic test we determine the best parameters for differential-time measurements and estimate their uncertainties, specifically for each waveform. We design a relocation strategy to work without a predefined velocity model, by formulating and inverting the problem to seek changes in both location and slowness, thus accounting for azimuth, take-off angles and velocity deviations from a 1D model. We solve the inversion explicitly in order to propagate data errors through the calculation. With this approach we are able to resolve a source volume few tens of meters wide on horizontal directions and around 100 meters in depth. There is no suggestion that the hypocentres lie on a single fault plane and the depth distribution indicates that their source is unlikely to be related to glacial processes as the ice thickness is not expected to exceed few tens of meters in the source area.

# 1. Introduction

Earthquake multiplets consist of very similar waveforms, often exceeding cross-correlation coefficients of 0.8 (Geller & Mueller 1980; Frémont & Malone 1987). They are common in tectonic and volcanic areas worldwide and they are likely to be caused by earthquakes occurring very close to each other and generated by similar, non-destructive, source processes (Geller & Mueller 1980). Because they consist of closely-spaced earthquakes, it is possible to determine relative relocation of the hypocentres with high accuracy (Poupinet *et al.* 1984; Fréchet 1985; Frémont & Malone 1987; Got *et al.* 1994; Slunga et al. 1995; Waldhauser & Ellsworth 2000; Thelen *et al.* 2008). The relative relocation method is based on the idea that closely-spaced events recorded at a common station will share similar path effects and site effects. If the hypocentral separation between two events is small compared to the station-hypocentre distance and scale length of velocity heterogeneities, and if the latter is big compared to the dominant wavelength of the waveforms, then the ray paths to a common station are similar and the relative time lag between the two events will depend on their spatial offset in the direction of the station (Waldhauser & Ellsworth 2000; Wolfe 2002).

Moreover, the location precision is improved by using high-precision waveform cross-correlation methods to determine the relative time measurements. This can be done either in the frequency domain (Poupinet *et al.* 1984) or in the time domain (Deichmann & Garcia-Fernandez 1992). The accuracy of the arrival-time differences between pairs of similar events is reported to be on the order of 0.001 s for micro-earthquakes recorded by local networks (e.g. Frémont & Malone 1987). This makes it possible to calculate the relative location between hypocentres with uncertainty on the order of a few meters to tens of meters (Waldhauser & Ellsworth 2000).

This is particularly useful at volcanoes, where earthquakes are often characterized by unclear phase onsets and their arrival-time determination can be highly imprecise with manual phase-picking. The relative location of earthquake multiplets is, therefore, a common practice at volcanoes worldwide. Got *et al.* (1994) relocated 250 earthquakes beneath Kïlauea that defined a nearly horizontal plane of seismicity at 8-km depth. Rowe *et al.* (2004) relocated approximately 17,000 similar earthquakes on Soufrière Hills volcano, Montserrat. On Mount Pinatubo, Philippines, Thelen *et al.* (2008) relocated several multiplets associated with the 2004-2006 eruptive sequence at Mount St. Helens and suggested that they were related to pressurization of the conduit system.

Two techniques are commonly used for relative relocation of earthquakes. One is the master-event approach, where all other events are relocated with respect to one, the master event (Ito 1985; Scherbaum & Wendler 1986; Frémont & Malone 1987; Van Decar & Crosson 1990; Deichmann & Garcia-Fernandez 1992; Lees 1998). Alternatively, cross-correlation time delays can be computed for all possible event pairs and combined in a system of linear equations to determine hypocentroid separations (Got *et al.* 1994; Waldhauser & Ellsworth 2000). In addition to adding more constraints to the model parameters, this strategy makes it possible to increase the spatial extent of the cluster that can be relocated, as there is no need for all events to correlate with the master.

The ability of the relative-relocation technique to recover the relative locations well depends on i) the geometry of the network, ii) the accuracy of differential-time measurements, iii) the deviations from the assumption that the ray paths do not change within the cluster of events, iv) the direction of the rays leaving the source, depending on 3D velocity variations (Slunga *et al.* 1995; Michielini & Lomax 2004). The relative relocation problem is usually solved in a 1D velocity model (e.g. Waldhauser & Ellsworth 2000) or with a constant slowness vector for each station to the cluster (Got *et al.* 1994). However, in these approaches the source of error represented by the uncertainties in the ray directions in the source volume is not taken into account. Michielini & Lomax (2004) showed how the initial 1D velocity model used, determining the take-off angles, influences the resulting shape of the relocated cluster. Moreover, in highly heterogeneous media, such as in volcanic areas, strong lateral heterogeneities can cause considerable deviations in the direction of the seismic rays from the straight path assumed in a 1D velocity model. This, in turn, can affect the spatial direction in which the earthquake location is re-adjusted as constrained by each station's differential time.

We propose a relative relocation strategy that does not rely on a 1D velocity model, but rather seeks changes in slowness vectors together with changes in relative relocation of hypocenters. We apply the relative-relocation technique to a cluster of LP (Long Period, Chouet 2003) seismic events located on the south flank of the subglacial volcano Katla, in south Iceland. This seismicity started in July 2011, in association with an unrest event which culminated in a glacial flood. Seismic events in this part of Katla volcano had never been recorded before. Since they occurred in a glaciated area, they can be generated by either glacial or volcanic processes. Sgattoni *et al.* (2015) suggested that they are associated with hydrothermal processes. A closer insight into the relative relocation of the hypocentres can give a useful contribution to the source interpretation.

The extreme similarity of the waveforms indicates a very small spatial distribution of the hypocenters. Moreover, indications of strong path effects, together with the poorly known velocity model for the site, motivated the development of a specific strategy to optimize both the differential-time measurements and the relative relocation technique. We conduct both cross-correlation measurements and relative-relocation inversion so that the uncertainty is carefully estimated and the sources of error minimized in each step. We perform a statistical test to evaluate the best cross-correlation parameters and uncertainties of differential-time measurements specifically for each station and each seismic phase used. This alleviates the need to use generalized statistical assumptions about errors. We then relocate the events with a master-event relocation strategy, inverting for both changes in location and in slowness, in order to account for azimuth, take-off angles and velocity deviations from a 1D model. We solve the inversion explicitly in order to propagate data errors through the calculation. We also perform synthetic tests to evaluate the ability of this approach to recover relative locations and slowness-vector components.

## 2. Seismic data

### 2.1 Seismic network

Following the eruptions of the neighbouring Eyjafjallajökull volcano in 2010, the IMO augmented the seismic monitoring network around Katla from 5 to 9 stations. Moreover, 9 temporary stations were deployed by Uppsala University between July 2011 and August 2013.

Most of the stations were equipped with broadband sensors, 5 Guralp ESPA , 4 Guralp CMG3-ESPC and 1 Geotech KS-2000(LLC), all with flat response from 60 s to the Nyquist frequency (50 Hz). 5-second Lennartz populated the remaining 8 stations. Data were recorded and digitized with Guralp and Reftek systems at 100 sps. Stations were powered with batteries, wind generators and solar panels. All the instruments recorded in continuous mode, but some technical problems (e.g. power failure) mainly due to harsh weather conditions (especially in winter time), prevented some stations from working continuously during the whole operation time.

Because the seismic events we analyse are very small (magnitude lower than 1.2; Sgattoni *et al.* 2015), the signal to noise ratio (snr) is low at distant stations. Therefore, we used data from 13 out of 19 stations (Fig. 1).

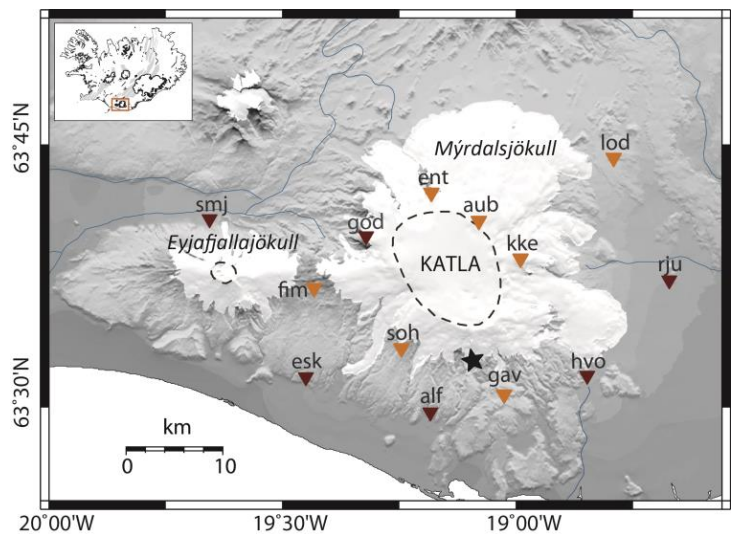

Fig. 1. Map of Mýrdalsjökull and Eyjafjallajökull showing the seismic network used in this study. Dark brown triangles: permanent IMO seismic stations. Orange triangles: temporary Uppsala University seismic stations operating between July 2011-August 2013. The star marks the new Gvendarfell cluster on the south flank. The Katla and Eyjafjallajökull caldera rims are outlined by dashed lines. White areas are glaciers.

## 2.2 LP seismic events, Katla south flank

The LP events recorded near Gvendarfell on Katla's south flank have been described in detail by Sgattoni *et al.* (2015). We report here the main features. The seismicity is shallow and located on the southern side of Mýrdalsjökull glacier. It is characterized by small magnitude (~ -0.5-1.2 $M_L$), long-period earthquakes with an emergent P wave and an unclear S wave (Fig. 2). The frequency content is narrow banded around 3 Hz at most stations (Fig. 2). All events have remarkably similar, nearly identical waveforms with correlation coefficient ≥ 0.9 at the nearest stations, throughout the whole time period investigated (March 2011 – August 2014). The size distribution is non-monotonic, with small events below magnitude $M_L$ = 0.2 and bigger events up to $M_L$ = 1.2.

The signals are characterized by a number of distinct seismic phases, whose nature is difficult to understand, as the waveforms are heavily contaminated by secondary phases generated by strong path effects. It is in general possible to recognise a P phase and a secondary wave package whose interpretation is not clear, probably containing both S waves and surface waves. Although unclear, we will refer to it as S wave (Fig. 2).

Around 1800 events have been detected with cross-correlation of a sample waveform with continuous data between July 2011 and August 2013. The temporal evolution shows striking features: a regular time pattern with 6 events per day at 4 hour intervals began a few hours before the tremor burst of the 2011 unrest episode, which occurred on July 8th-9th. A seasonal variation in the event rate is also observed, with maximal activity in late summer 2011, 2012 and 2013.

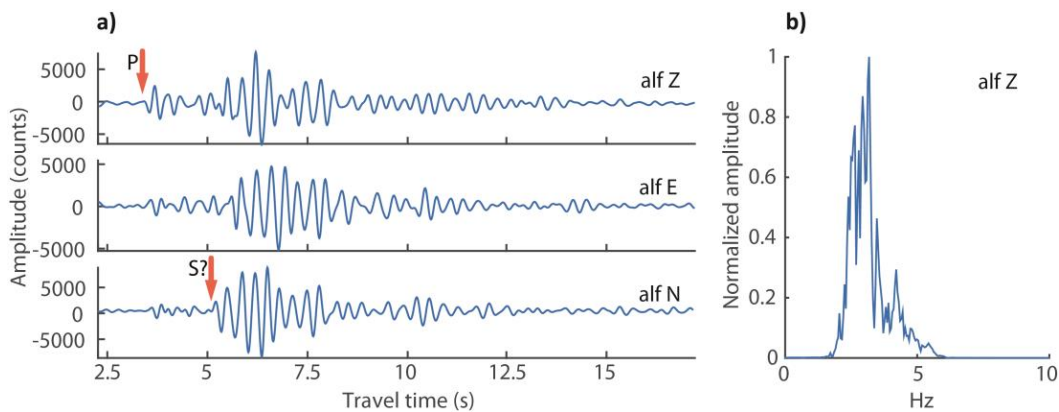

Fig. 2. a) Example seismograms of a Gvendarfell event at stations ALF. The amplitude unit is digital counts, proportional to velocity. b) Normalised amplitude spectra of the Z component of the same event at stations ALF.

## 3. Differential time measurements

As all the waveforms are extremely well correlated, with cross-correlation coefficients higher than 0.9 (mostly ≥ 0.95; Fig. 4), at the nearest stations, we expect a very small spatial distribution of the hypocentres. We can crudely estimate the maximum size of the source distribution requiring a phase difference of less than a fraction of a period in order to achieve such high levels of correlation. Taking this fraction to be a quarter and assuming an average velocity of 3.5 km/s and a dominant frequency of 3.5 Hz, we constrain the source region to be smaller than 250 m in size. Therefore, in all steps leading to the relative relocation results, we attempt to minimize the uncertainty and also carefully estimate it, from the differential-time measurements to the resulting relative locations. In order to measure the relative times as accurately as possible, we set up a synthetic test to identify the best parameters to use for the cross-correlation. We then use the synthetic test to evaluate their uncertainty, later used as weights for the relative relocation inversion.

### 3.1. Statistical test for cross-correlation

We perform a synthetic test that measures the time shifts between a clean waveform (the template waveform) and the same waveform with different levels of random noise added. The random noise is generated as white noise and then filtered with the same filter as the template and adjusted in amplitude to constitute the specific snr, which is defined as the ratio between the rms signal amplitude in the correlation time-window and rms noise amplitude in a window of equal length before the P arrival.

At each station and for each component of the signal:
- P and S phases are identified and extracted from the template waveform, previously band-pass filtered between 2-4, 1-5 or 2-5 Hz depending on the station (the S phase is identified only at stations ALF, GAV, SOH, HVO, ESK, FIM and RJU).
- The extracted P/S window is tapered with a 10% cosine taper.
- A template P/S window is correlated in the time domain with the same window with noise added (without applying any time shift so that the differential time is known to be exactly zero). A parabolic interpolation around the peak of the correlation function is done to estimate the time shift with sub-sample precision.

The test is repeated for different snr (signal to noise ratios), varying from 1 to 10, and different widths of the P and S windows, from 0.5 s to a few seconds.

The whole process is performed at least 100 times, each time generating a new random noise vector (Fig.3). The std (standard deviation) of the calculated time shifts is then computed and its behaviour is analysed to determine, for each station, i) the best window width to use for each type of wave and component at each station, ii) the expected std as a function of the correlation coefficient. The first is done as follows:

- The correlation coefficients obtained for different snr and window widths are plotted against the window width used, always starting at the P arrival time at a given station.
- Two peaks are identified in this plot. One peak occurs before the S arrival and represents the best window length for the P wave. The other peak occurs after the S arrival and corresponds to the sum of P and S windows that have best correlation. The best window width for the S wave, therefore, corresponds to the time of the second peak minus the S-P time at a given station.

Fig. 3 shows an example for station ALF: a first peak of correlation is observed at 1 sec and this is, therefore, the width chosen for the P wave. At this station the S phase arrives 1.6 seconds after the P phase and this is reflected by an increase in correlation coefficient starting after this time and reaching a new peak around 4 seconds. Therefore, the window width chosen for the S wave is 2.4 seconds.

The analytical std as a function of the correlation coefficient is then estimated for the chosen window length (Fig.3):
- The standard deviation obtained for different snr and window lengths is plotted as a function of correlation coefficient.
- An empirical curve is fitted to the data and later used to estimate the uncertainty of the differential-time measurements of the real data, based on the correlation coefficient.

The same procedure is repeated for P and S phases and all components. The uncertainty of the differential time estimates is in most cases lower than the sampling interval (0.01 s) and as low as 1 ms (Fig.4).

This synthetic test allows for optimization of the differential-time measurements in order to minimize uncertainty. It also allows us to estimate uncertainty specifically for each waveform, thus avoiding further generalized statistical assumptions about errors.

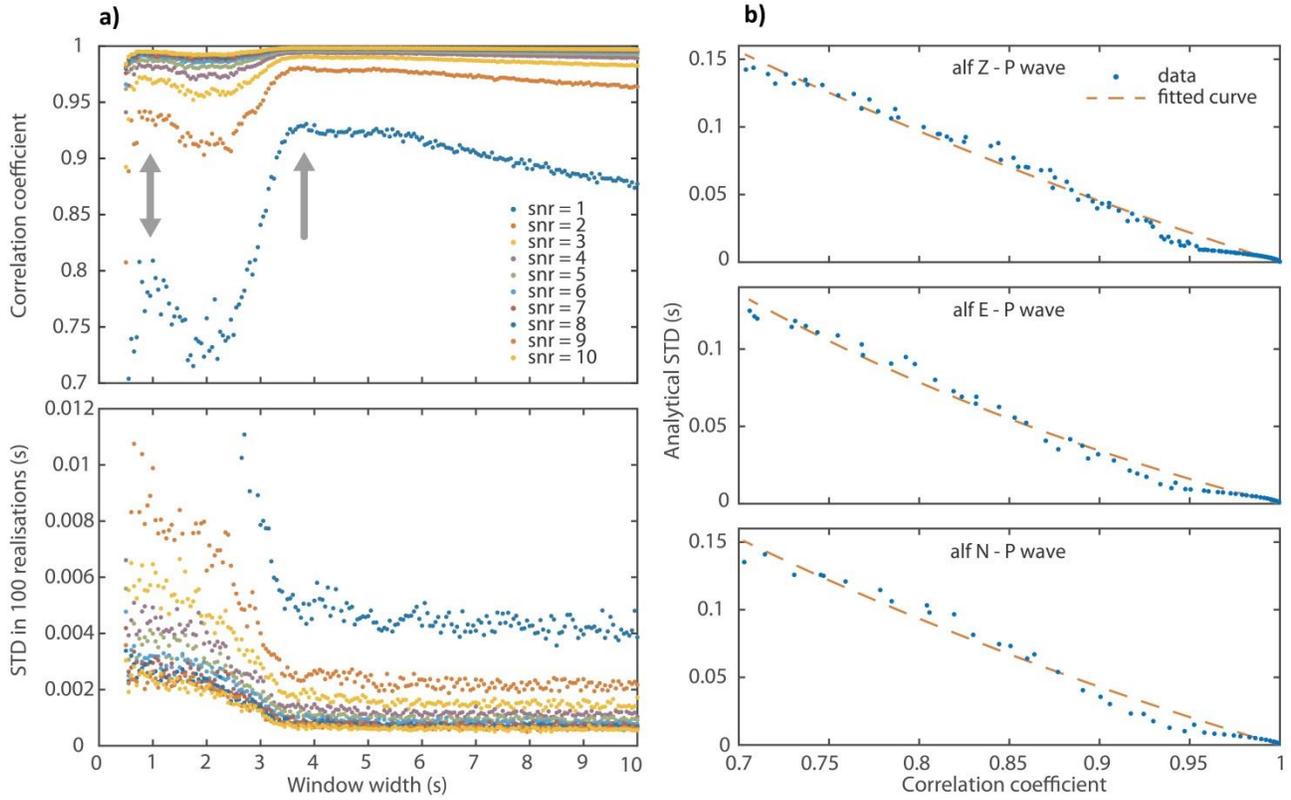

Fig. 3. Statistical simulation to determine uncertainty of cross-correlation measurements. Data from station ALF. a) correlation coefficient and std obtained for different widths of the window used for correlation and different values of snr. Window widths are measured starting from P arrival time. A first peak is observed at around 1 sec (double arrow): this is the width chosen for P wave. At station ALF the S-P time is 1.6 s and this determined increase in correlation after this time and reaching a new peak at ~4 seconds (single arrow). The window width chosen for S wave is therefore 2.4 s (corresponding to 4 s minus 1.6 s). As expected, uncertainty decreases with increasing snr. b) analytical standard deviation as a function of correlation coefficient, measured for chosen window width (1 sec) and varying snr. Results for 3 components of P wave at ALF. An empirical fitted curve is then used to estimate uncertainty of differential-time measurements between the template event and all the others.

### 3.2. Cross-correlation measurements

Once the best parameters for the time-difference estimation are determined, a cross-correlation scheme is built to correlate a reference event (later used as master event for the relative relocation) with all other events (P and S phases separately), at all stations, for all components. Sub-sample estimates of time lags are achieved in the time domain through polynomial interpolation of the cross-correlation coefficient peak. Since not all stations have been working at the same time for the entire period of study, it is not possible to identify one unique reference event. An event which occurred on Oct. 10th 2011 is chosen for stations ALF, GAV, KKE, HVO, AUB, ENT, RJU, ESK, FIM, SLY, SMJ. An event on Feb 18th 2013 is chosen for SOH and on Oct 1st 2011 for LOD.

For each station, the differential-arrival times and the corresponding uncertainties are estimated for P and S waves for all 3 components. Moreover, for each phase (P and S) the weighted average of the time shifts for the different components by their uncertainty is computed. Ultimately, the best estimates (in terms of low error) are selected between the 4 values obtained (3 components and weighted average) and the uncertainty used to weight the data in the relative-relocation inversion.

The cross-correlation times are selected by setting a lower threshold for the cross-correlation coefficient as high as 0.9 for the closest stations and 0.8 for more distant stations and only event pairs with at least 6 time measurements are used (Fig. 4). We also discard some outlying data, with uncertainty greater than 0.03 s.

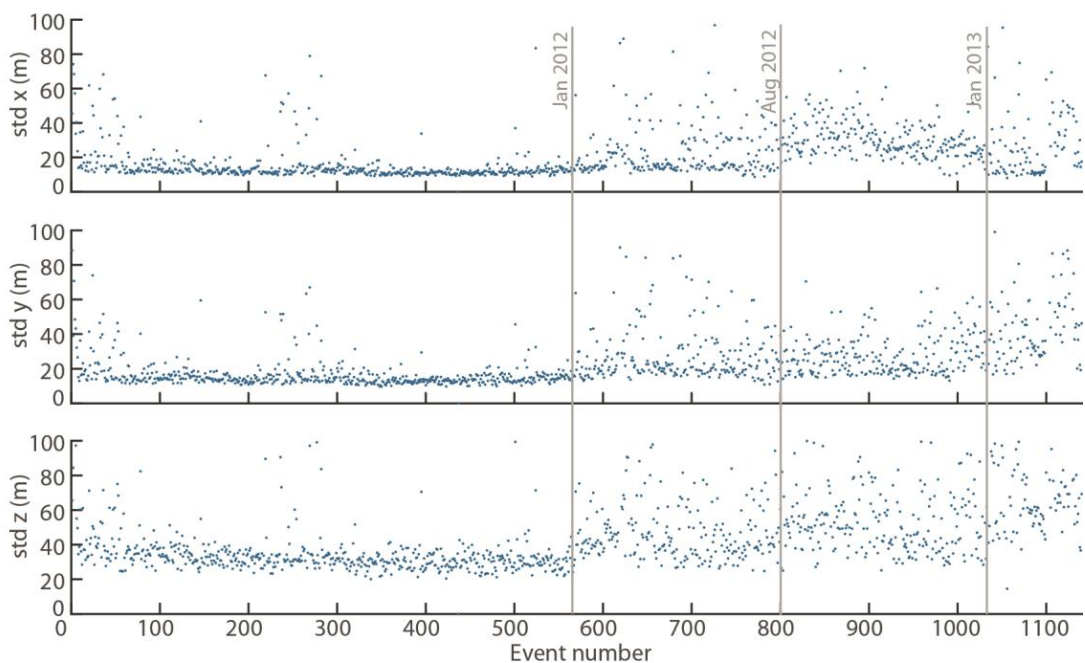

Fig. 4. Correlation coefficient (a), analytical error std (b) and number of observations for each event used in the relative relocation (c).

This reduces the number of relocated events to 1140. This is mainly due to the small magnitude of the events, in particular the smaller-magnitude group, which is only observed at a few nearby stations. From around August 2012 a decrease in average magnitude from ~1 $M_L$ to ~0.5 $M_L$ is also observed for the larger-magnitude group of events (Sgattoni *et al.* 2015). There is a greater loss of data in the second part of the dataset, after January 2012, when a slight decrease of correlation coefficient occurs and some data are lost due to technical problems at some seismic stations (Fig. 4). The decreased correlation coefficient causes increased uncertainty of differential-time measurements (Fig. 4), which starts in January 2012, increases until August 2012 and remains fairly stable after that. This pattern does not clearly correlate with changes in the seismic network configuration.

## 4. Relative-relocation method

As no catalogue locations exist for all 1200 events, a master event strategy is used to relocate the events starting from the same initial location and relocating them with respect to the fixed location of a master event (Ito 1985; Scherbaum & Wendler 1986; Frémont & Malone 1987; Van Decar & Crosson 1990; Deichmann & Garcia-Fernandez 1992; Lees 1998). An event which occurred on October 10th 2011, used as reference for cross-correlation time measurements at most of the stations, is chosen as a master event. As location coordinates, we use the highest-probability hypocentral coordinates from the combined probability density of non-linear absolute locations obtained by Sgattoni *et al.* (2015), at N63º32.772' and W19º05.988' and depth is referred to the average elevation in the source region (at 800 m.a.s.l.). Since a different reference event was used for cross-correlation measurements at stations SOH and LOD, we made sure that the three reference events used had at least 8 differential time information linking them.

In the first instance, the routine is built to relocate the events in a 1D velocity model, similar to most relative-relocation strategies (e.g. Slunga *et al.* 1995; Waldhauser & Ellsworth 2000). However, the data misfit achieved is not satisfactory, as the data are not explained by the model locations close enough to their level of uncertainty (known from the synthetic test explained above). The misfit, normalized by the data covariance, exceeds the expectation of the chi-squared distribution by a factor of 5.

We think this is partially due to effects of lateral heterogeneities that are expected to be strong in the crust in the area and may cause considerable deviations in the direction of the seismic rays from the straight path assumed in a 1D velocity model. Since this direction can vary considerably because of the local heterogeneities, this introduces inconsistencies that cannot be explained by the model. In addition, the correlation-time measurements are integrated measurements of wave packages extending over 0.5 to several seconds. They are likely to contain waves of varying type and geometry due to scattering. This is confirmed by their complex particle motion. Therefore, the effective slowness of these waves as they leave their source may differ from that predicted by a simple 1D model, in terms of azimuth, incidence angle and effective local velocity. Thus, the strategy is adjusted to account for this and the slowness vectors (3 spatial components per station per wave type) are included as model parameters in the inversion. Azimuths, incident angles and P/S velocity are allowed to be modified by the inversion, in order to account for velocity heterogeneities.

### 4.1. Formulation of the problem

At a given station, the arrival time, $t$, for an earthquake, $i$, corresponds to the sum of the origin time, $\tau$, and the travel time, $T$, as a function of the event spatial coordinates, $x_i$:

$$t_i = \tau_i + T(x_i) \tag{1}$$

We assume the locations are around a point, $x_0$, with only small changes, $\delta x_i$, much smaller than the propagation distance:

$$x_i = x_0 + \delta x_i \qquad |\delta x_i| \ll propagation\ distance \tag{2}$$

If the distances between the events are much smaller than the propagation distance from source to receiver, then the differences in path from the events to the same station can be described as planar. In our example, the distances between source and receiver are in the range of 6-30 km and the events are likely to be generated at distances from the centre of the cloud on the order of ≤ 100-150 meters, as apparent from the non-linear absolute locations presented by Sgattoni *et al.* (2015) and from the extremely similar waveforms. The events are so close to each other that a first order, linear or planar approximation of the travel-time function is justified. Therefore, we can apply a linear approximation:

$$T(x_i) = T(x_0 + \delta x_i) \approx T(x_0) + u \cdot \delta x_i \tag{3}$$

where $u$ are linear coefficients with the unit of slowness (sec/km). Consequently, the differential arrival-time, $\delta t$, between events $i$ and $j$ at a given station, can be expressed as:

$$\delta t_{ij} = t_i - t_j = \tau_i - \tau_j + u(\delta x_i - \delta x_j) \tag{4}$$

This is similar to the formulation by Got *et al.* (1994) who used a constant slowness vector for each station to the cluster. Instead, we allow the slowness vector to each station to vary in the inversion. Equation (4) is non-linear in the last term when both **u** and $\delta x_i$ are unknown and the model parameters are coupled, as they appear as a product. So, we linearize again by differentiating with respect to slowness and location parameters, to seek changes in both slowness and location:

$$\delta t_{ij} \approx \tau_i|_0 - \tau_j|_0 + d\tau_i - d\tau_j + \boldsymbol{u}_0(\delta x_i - \delta x_j) + \boldsymbol{u}_0(d\delta x_i - d\delta x_j) + d\boldsymbol{u}(\delta x_i - \delta x_j)|_0 \quad (5)$$

where the 0 subscript indicates the initial guess (or estimate at previous iteration) and d indicates the change of the model parameter. In each iteration we solve for perturbation of location parameters (spatial and origin time) and perturbation of the slowness vector, thus obtaining an update of location and slowness. We also apply constraints to the slowness vector within reasonable bounds in order to avoid absurd geometry configurations such as rays leaving the source in opposite direction with respect to the station location.

The initial slowness vector **u** is determined from azimuth angles, $\alpha$, and incidence angles, $\varphi$, for each station $j$, (estimated in a 1D velocity model) and constant initial velocities, $v$, for P/ S waves:

$$\boldsymbol{u}_j = \left(-sin\alpha_j \, sin\varphi_j \frac{1}{v}; \, -cos\alpha_j \, sin\varphi_j \frac{1}{v}; \, -cos\varphi_j \frac{1}{v}\right) \quad (6)$$

The initial P velocity is set as the P velocity at the master event hypocentral depth, corresponding to 3.5 km/s. The S velocity is set as the P velocity scaled by a factor of $1/\sqrt{3}$. In total, the slowness parameters are six per station (3 for P and 3 for S). We do not constrain the angles to be the same for P and S, as there can be different scattering phenomena with different influence on P and S azimuth and incidence angles.

### 4.2. Inversion

As we do not have catalogue locations for all events, before inverting for relative relocations, we need an estimate of origin times. So, we first formulate the problem in order to invert for origin time. This is a linear problem in which the differential arrival times correspond to the sum of the differential origin times and differential travel times:

$$\delta t_{ij} = \tau_i - \tau_j + T(\boldsymbol{x}_i)|_0 - T(\boldsymbol{x}_j)|_0 \quad (7)$$

We solve the problem by assuming that all events are located at the master event location (i.e. all relative locations equal to zero). Therefore, the travel times for all events are the same and equation (7) simplifies to a difference of origin times.

We combine all data in a system of linear equations of the form:

$$\boldsymbol{WGm} = \boldsymbol{Wd} \tag{8}$$

where $\boldsymbol{G}$ is a matrix of size N x Nev (N is the number of differential-time measurements; Nev is the number of events), $\boldsymbol{m}$ is the model vector of length Nev, containing origin times, $\boldsymbol{d}$ is the data vector of length N, and $\boldsymbol{W}$ is a diagonal matrix containing weights.

We then add hard constraints in the form of Lagrange multipliers in order to fix the master event origin time to a reference time (that we set to 0). The problem is overdetermined and can be solved in a weighted least-squares sense, where the weights are set as the inverse of the data covariances:

$$\boldsymbol{m} = [\boldsymbol{G}^T \boldsymbol{C}_d^{-1} \boldsymbol{G}]^{-1} \boldsymbol{G}^T \boldsymbol{C}_d^{-1} \boldsymbol{d} \tag{9}$$

where $\boldsymbol{C}_d$ is the data covariance matrix, a diagonal matrix containing data variances determined by the synthetic tests described in the previous section. In order to determine the inverse of the product matrix $[\boldsymbol{G}^T \boldsymbol{C}_d^{-1} \boldsymbol{G}]$, we use the singular-value decomposition (SVD) method. The system is well-conditioned and there is no need to regularize.

After solving for the origin time, an iterative process is set up to invert equation (5) alternatively for relative changes in location, $(d\delta x_i - d\delta x_j)$, and changes in slowness, $d\boldsymbol{u}$. The inversion strategy is the same as before, but the size of the matrices and the hard constraints change. In the inversion for location parameters, the model vector $\boldsymbol{m}$ has length 4Nev ($d\delta x_1$, $d\delta x_2, d\delta x_3, d\tau$ ) and the size of $\boldsymbol{G}$ is N x 4Nev. The Lagrange multipliers consist in this case of four additional equations to constrain all changes of hypocentral parameters of the master event to be zero. Also in this case, the system is well-conditioned as we pre-filtered the data so that all event pairs have at least 6 observations. This way all events are well linked to each other.

When inverting for slowness perturbations, the size of the matrices decreases significantly as the number of model parameters reduces to the size of the slowness vector, i.e. 6 times the number of stations (Nst), and $\boldsymbol{G}$ is therefore N x 6Nst. Again, the inversion strategy is the same and the matrix is inverted with SVD. In this case, some regularization is needed to exclude the zero eigenvalues originating from stations that have no information for some slowness components, depending on which phases have been used for the cross-correlation. Also, some small eigenvalues can occur if only little information is used for some stations and if, due to the geometry of the problem, some directions are poorly constrained. A threshold value for eigenvalues is found by trial and error.

After the inversion is performed, the resulting perturbations of slowness are checked in order to apply constraints to incidence angle, azimuth and velocity variations, so that they do not exceed specified values. P and S wave velocities are allowed to vary within ± 1 km/s from the initial values. The azimuth angles can change within ± 30 degrees from the initial direction and the incidence angles within ± 20 degrees.

The inversion scheme is therefore the following:
- Iteration 0: inversion for origin time and inversion for location parameters (using the initial slowness vector)
- Iteration 1…n: inversion for slowness and inversion for location parameters (using updated slowness vector).

The process is iterated until the misfit reduction is negligible, for a total of 7 iterations.

## 4.3. Model covariance estimation

The uncertainty of the location model parameters is then determined by propagating the estimated data covariances to the model parameters. However, this is sufficient only if the data are appropriately explained (within their uncertainty). In our case study this does not happen, as the final misfit achieved is bigger than the value expected statistically. If data are appropriately explained, the data misfit scaled by data covariance, $Q$, is expected to equal the number of degrees of freedom of the model, if the model errors are Gaussian:

$$E\{Q\} = e^T C_d^{-1} e = n - r \qquad (10)$$

where $E\{Q\}$ is the expectation of $Q$, $e$ is the prediction error, $n$ is the number of observations and $r$ is the number of eigenvalues used in the inversion (the degrees of freedom in the model). In our case the misfit is bigger than $(n - r)$. This means that errors in the problem are not fully described by measurement errors. Additional errors occur, possibly due to simplification of the theory or limited knowledge about the velocity model. We account for this by adding a uniform diagonal covariance to our estimated data covariance matrix, so that the expected misfit indicates that data are appropriately explained. We do this by calculating $Q$ as a function of data variance, $\sigma_d^2$, with a constant, c, added to the diagonal elements:

$$\hat{Q} = e^T(C_d^{-1} + cI)^{-1}e \tag{11}$$

where $I$ is the identity matrix. We solve $\hat{Q} = n - r$ for $c$. We therefore obtain a new data-covariance matrix, which includes both measurement errors and errors due to simplifications of the way the forward problem is described and solved. We use this as a new estimate of the total data covariance and propagate it through the calculation, to estimate the model covariance, i.e. the uncertainty of relative relocation.

This strategy of adding a random component to the data variance estimates of measurement errors assumes that the errors in the data are independent of errors due to the simplification of the forward problem. Another possible strategy would be to scale the errors up according to the residual misfit, but we have no reason to expect that the unaccounted for errors are correlated with and proportional to the analytical measurement errors.

The resulting uncertainty of spatial location parameters is on the order of 15-20 meters for the horizontal components and 30-40 meters for the vertical. A general increase in uncertainty is observed from January 2012, correlating with bigger data uncertainties (see Section 3.1) and associated with less phases available per event-pair, as explained in Section 3.2.

## 5. Synthetic tests

In order to test the behaviour of the inversion and the ability of the program to recover both hypocentre locations and slowness components, we performed several synthetic tests with the station configuration and geometry of the problem of our case study on the south flank of Katla. Using equation (4), we generate differential times for all event pairs of a set of 50 events with random hypocentral locations within a 300x300x300 m volume. The data are generated using an initial slowness vector for the same 1D model that was used to construct the slowness for the inversion of real data (Section 4.1).

We generate perfect data and a slowness vector perturbed with random Gaussian errors. We track, iteration by iteration, the slowness vector std with respect to the true slowness and the spatial and temporal mislocations of the hypocentres. We repeat the test at least 100 times and compute an ensemble average of the results for all realizations. We repeat the same process for different size initial slowness perturbations. We set the bounds of the constraints imposed during the inversion (Section 4.2) as the maximum perturbation allowed, corresponding to two std of the Gaussian distribution used to generate the random errors. We then perform several tests with decreasing percentages of this maximum perturbation, from 100% to 0.1%. In all tests data are fitted perfectly, while hypocentre locations and slowness components are recovered to some extent, depending on the initial slowness perturbation. The average results for 100 repetitions of this test show that:

- the smaller the perturbation of the initial slowness vector the better hypocentre locations are recovered (with perfect recovery for an unperturbed initial slowness vector). For small perturbations, most of the relocations occur at the initial, $0^{th}$ iteration. This is expected since in this case the non-linearity is weak;
- the slowness vector (in terms of azimuths, incidence angles and P and S velocity) is adjusted, iteration by iteration, towards the truth, with a reduction of the std (compared to the true slowness) of 50 to 70% achieved (greater proportional reduction when starting with less perturbed slowness;
- when the slowness perturbation applied is small (up to 20% of the maximum perturbation allowed) the hypocentral locations do not change significantly after the initial $0^{th}$ iteration, although the slowness vector changes and moves towards the truth;
- for larger slowness perturbations (up to the maximum), the hypocentral mislocations reduce with iteration, but the mean mislocation reduction (in terms of distance between true and calculated locations) only reaches a maximum on the order of 30-40% for the maximum initial slowness perturbation.

These tests demonstrate effects of the non-linearity of the problem and that trade-offs occur between location parameters and the slowness vector. The function we try to minimize has multiple minima that may prevent the inversion from reaching the global minimum. This, in turn, means that the final relative locations obtained may depend on the initial slowness vector. However, the true slowness vector is successfully recovered by at least 50% in all of our tests and, even for the larger initial slowness perturbations, the hypocentre mislocations are significantly reduced also, after the $0^{th}$ iteration, when introducing the inversion for slowness.

## 6. Relative relocation results

Since the synthetic tests demonstrate a trade-off between relative relocations and slowness vector, we compute the inversion with different starting slowness vectors and compare the results. We perturb the initial guess of the slowness vector (obtained as described in Section 4.1) with 25%, 50% and 100% of the maximum allowable random perturbation, for a total of 8 inversions. All inversions converge to similar results both in terms of relative relocations and slowness vectors. The initial misfit (after inversion for origin time), normalized by the data covariance and scaled by the number of degrees of freedom, is 9.2. At the $0^{th}$ iteration, it ranges for the 8 inversions between 4.9 and 5.5, with larger values for larger slowness perturbations. After 7 iterations, the misfit is reduced to values between 2.8- 3.1.

All inversions converge to a similar size and shape of the cloud of hypocentres and the slowness vector components move in the same direction. However, they do not converge to the same values, as observed in Fig.5 where the resulting azimuth angles are reported as average over the 8 inversions and corresponding std. The results indicate that the variation increases with distance between station and seismic cluster. This is expected, as scattering effects are likely to increase for longer travel-distance, together with the width of the Fresnel zone. In some cases, the P and S azimuth angles move to opposite directions compared to the initial value. Also, for P waves the azimuths deviate more from straight paths, compared to S waves.

The resulting relative relocations are shown in Figs. 6-7. Fig. 6 shows a comparison of the absolute IMO catalogue locations, absolute non-linear locations obtained by Sgattoni *et al.* (2015) and the relative relocations obtained in this study. The spatial distribution of 870 catalogue locations spans an area several km wide, with a formal uncertainty on the order of 1 km. The non-linear locations of 32 events, obtained with the addition of 2 temporary stations within 2 km from the cluster, are concentrated in a smaller area, less than 1 km wide, with uncertainty estimates around 400 m. Our relative locations of 1140 events cover an even smaller area, few hundred meters wide, with horizontal relative uncertainty on the order of 15-40 m (Figs. 7-8).

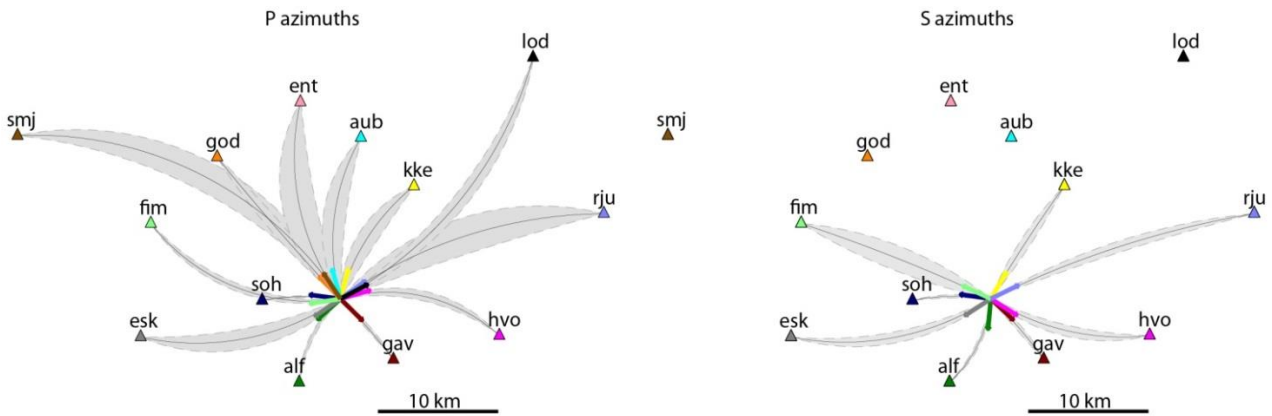

Fig. 5. Average azimuth vectors and their std for all 8 inversions performed, with different starting slowness vectors. P and S azimuths are drawn separately, as the inversion is performed separately.

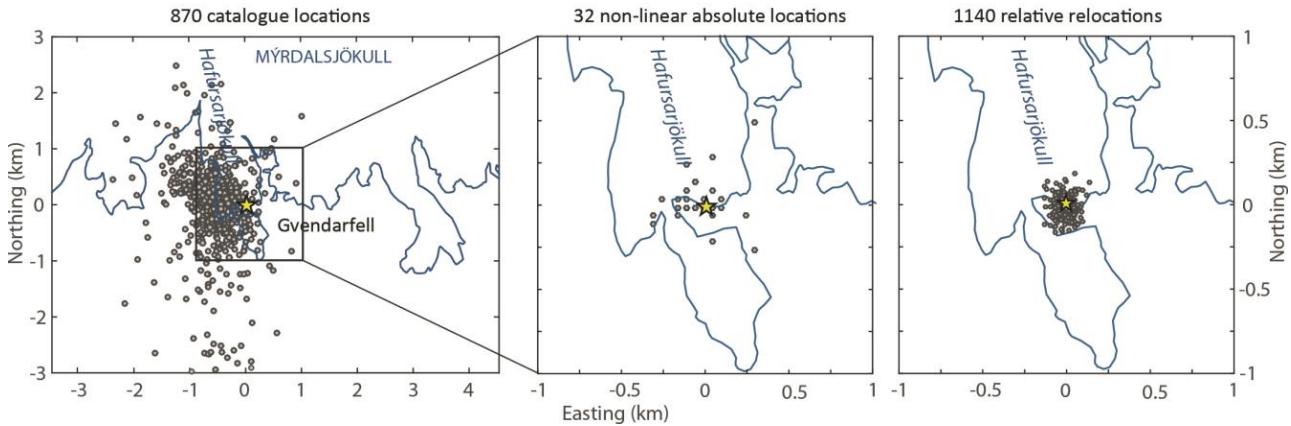

Fig. 6. Map view of 870 events from the IMO catalogue, 32 non-linear absolute locations (Sgattoni *et al.* 2015) and 1140 relative relocations. The yellow star is the master event location, corresponding to the centre of the non-linear absolute locations (at N63º32.772', W19º05.988' and depth corresponding to the local surface elevation) and is the origin of the axes scale. The blue line is the glacier outline derived from LiDAR DEM obtained in 2010 (Jóhannesson *et al.* 2013). The spatial distribution of the hypocentres is reduced with the relative relocation from several km to tens of m.

The std estimates for the 3 spatial components (Fig. 8) indicate that the uncertainty in depth is about twice that in the horizontal directions. There are also changes with time: starting from January 2012 the std increases sharply in all directions and its variability increases. This correlates with changes in correlation coefficient and related data uncertainty estimates and the decrease in the number of observations per event (Fig. 4). In Fig. 7 we report the locations and error bars for all events (950) with a smaller std than 60 m on all three directions. The average std on the horizontal components corresponds to 14 m before January 2012 and 33 m after that. The average std in depth increases from 32 m to 45 m. In order to estimate the size of the cluster, we derive the combined probability density distribution of the whole cluster by summing the distribution of all individual events (based on their location uncertainties). Although visually the size of the cluster appears to increase in the second time period (after January 2012, Fig. 7), there is in fact no significant change, since the uncertainty increases as well. The resolved size of the cluster is estimated to be on the order of 25x50x100 m (easting, northing, depth). There is indication of a shift in location between the two time periods on the order of 30 m towards south, but this is not resolved as it is below the level of uncertainty.

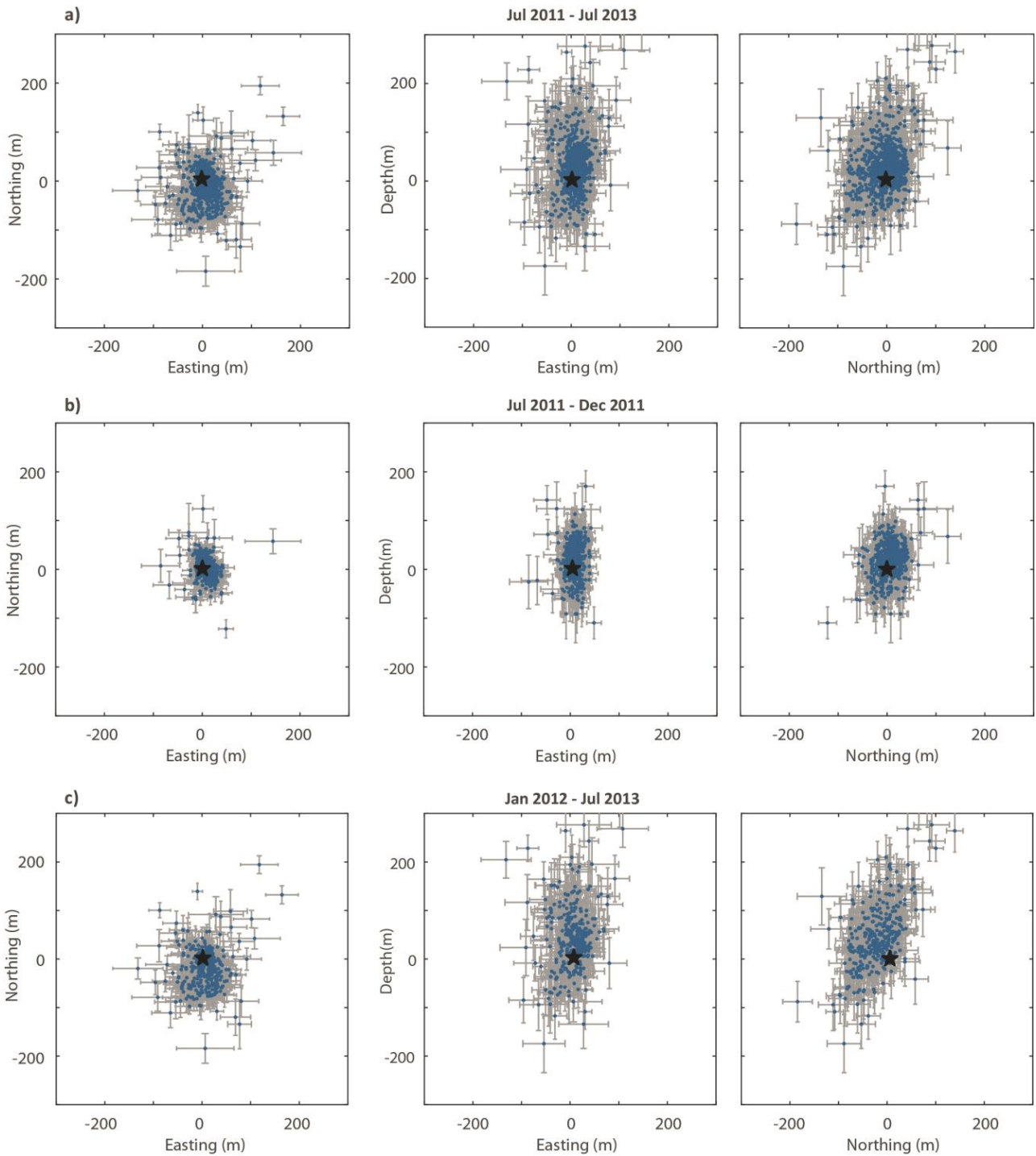

Fig. 7. Relative relocation results for all events with a smaller std than 60 m in all directions. Blue points are the locations and grey lines represent uncertainty (± std). The star is the master event location, corresponding to the origin point of the axes. a) 950 events for the entire time period (July 2011 and July 2013). b) 550 events, until December 2011 c) 400 events, from January 2012.

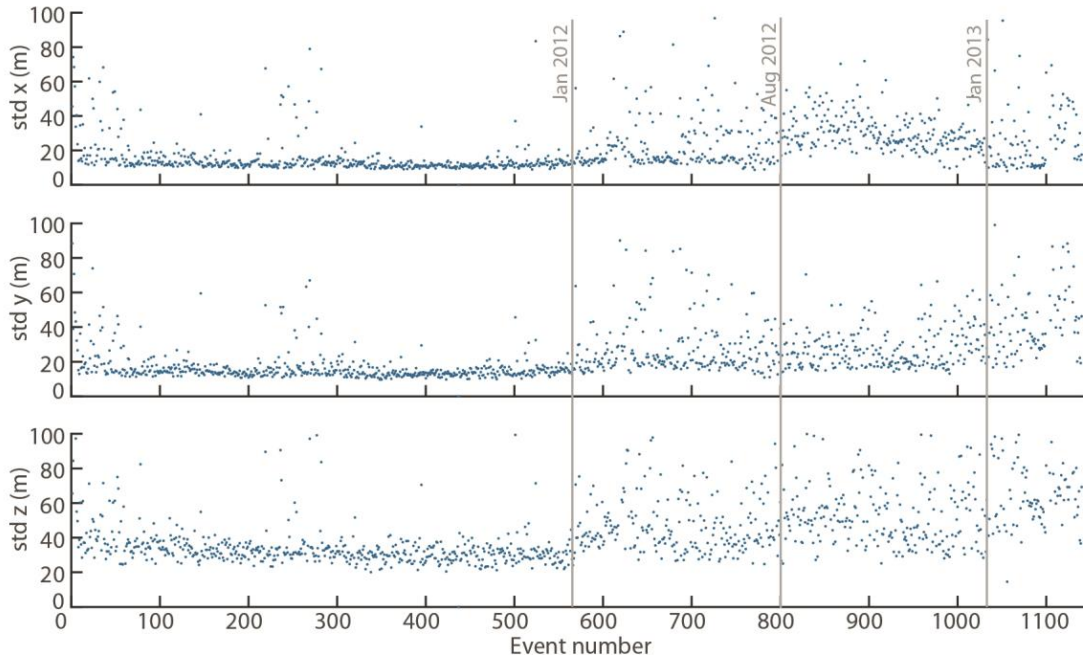

Fig. 8. Estimated std of the relative relocation spatial coordinates x (easting), y (northing), z (depth). A clear increase starts in January 2012 and peaks in August 2012.

## 7. Discussion and conclusions

We have located with a relative relocation strategy around 1100 shallow, repeating LP seismic events located on Katla volcano's south flank, at Gvendarfell. This seismicity started in 2011, in coincidence with an unrest episode that culminated in a glacial flood and is characterized by a strikingly regular temporal pattern, with regular intervals between repeating seismic events, modulated by a seasonal variation. Sgattoni *et al.* (2015) suggested that this seismic activity may be related to hydrothermal processes, although no evidence for hydrothermal activity was found in the area. As they occur at the rim of the glacier, both volcanic and glacial processes must be taken into account as possible sources. Information on the size and shape of the seismic cluster may help constraining the source process. This motivated our detailed study in order to extract information on the hypocentre distribution.

The extreme similarity of all waveforms indicates a very small spatial distribution of hypocentres. In order to extract meaningful information about size and shape of the cluster, we have optimized the cross-correlation measurements and relative-relocation process in order to minimize uncertainty. With a synthetic test we determined the best parameters for differential-time measurements and estimated their uncertainties, specifically for each waveform. This allowed avoiding further generalized statistical assumptions about errors.

We have then relocated the events with a master-event approach, which is justified by the small size of the problem: the waveforms are so similar that the master event is well correlated with all others. The 1D velocity model assumption that is usually made in most location strategies is not likely to describe reality in volcanic areas where strong heterogeneities are expected. Therefore, we have designed the relocation strategy to work without a predefined velocity model, by formulating and inverting the problem to seek changes in both location and slowness. This strategy accounts for azimuth, take-off angles and velocity deviations from a 1D model and allowed to considerably improve the data fit. When allowing the slowness vectors to be changed during the inversion, the misfit is reduced by almost 50% and approaches its expected value. In order to propagate data errors through the calculation, we have solved the inversion explicitly and estimated a location covariance matrix.

We have tested the program synthetically and observed a trade-off between relocations and slowness that lies in the nature of the problem, which is non-linear and in which the model parameters are coupled as factors in the same term (Eq. 4). For this reason, we have performed the inversion with several initial slowness vectors. All inversions resulted in similar hypocentre distributions and slowness values and angles.

The Gvendarfell seismic cluster appears to be distributed over a volume with depth distribution on the order of 100 m and horizontal distribution on the order of 25x50 m. This allows some considerations about the interpretation of the source:

- there is no suggestion that the shape of the cluster has a single plane-like geometry. Therefore, there is no evidence that the seismic events are generated by fault movement, despite the fact that a recent fault was identified in the area (Sgattoni *et al.,* in preparation).
- the depth distribution of the hypocentres suggests that these events are unlikely to be generated by glacial processes, as the ice thickness is not expected to exceed few tens of meters in the area where the cluster is located. Therefore, volcano-related processes, magmatic or hydrothermal, are more likely, as suggested also by Sgattoni *et al.* (2015).
- the size and shape of the cluster do not exclude or point to a specific volcano-related source. In the case of a hydrothermal source, the size may be consistent with e.g. a crack or a crack volume filled with hydrothermal fluid. Alternatively, a small batch of magma rising at shallow depth may act as a source. The size of the cluster is consistent with the size of silicic magma bodies identified in the Gvendarfell area (Sgattoni *et al.,* in preparation).

Some indications of minor temporal changes are suggested by decreased cross-correlation coefficient and increased location uncertainty after January 2012. While the location uncertainty may be influenced by e.g. network configuration changes, a systematic decrease in correlation coefficient may be associated with a decrease in the size of the events or with changes in either source process or hypocentre locations. There is no clear correlation between magnitude and correlation coefficient variations or between the systematic increase in location uncertainty and the changes in the network configuration. We suggest, therefore, that time changes in either the source process or hypocentre location may have occurred starting from January 2012. There is an indication of a shift of the hypocentres towards south, but this is below the uncertainty level. It is not straightforward to infer what this time change would imply for the source interpretation.

# Aknowledgements

The authors would like to thank the Icelandic Met Office for access to waveform of the Gvendarfell events. The temporary deployments producing data for this study were supported by CNDS (Centre for Natural Disaster Science, www.cnds.se) at Uppsala University and the Volcano Anatomy project, financed by the Icelandic Science Foundation. This work was funded by the University of Bologna, University of Iceland and Uppsala University, as a part of a joint PhD project.